\documentclass{article}

\input{tcilatex}

\begin{document}

\begin{center}
{\LARGE Theory of decoherence in Bose-Einstein condensate interferometry}

\bigskip

\textbf{B J Dalton}

ARC Centre for Quantum-Atom Optics

and Centre for Atom Optics and Ultrafast Spectroscopy

Swinburne University of Technology

Melbourne, Victoria 3122, Australia

Email: bdalton@swin.edu.au

\bigskip 
\end{center}

\textbf{Abstract. }A full treatment of decoherence and dephasing effects in
BEC interferometry has been developed based on using quantum correlation
functions for treating interferometric effects. The BEC is described via a
phase space distribution functional of the Wigner type for the condensate
modes and the positive P type for the non-condensate modes. Ito equations
for stochastic condensate and non-condensate field functions replace the
functional Fokker-Planck equation for the distribution functional and
stochastic averages of field function products determine the quantum
correlation functions.

\section{Introduction}

Bose-Einstein condensates (BEC) in cold atomic gases are an example of a
quantum system on a macroscopic scale. Well below the transition temperature
essentially all bosons occupy small number of single particle states (or
modes) -- in simple situations only one mode. Interferometry based on BECs
(such as by splitting a BEC in a single trap into two traps and then
allowing the BECs to recombine) offers possible improvements in precision
over single atom interferometry by a factor given by square root of the
boson number $N$ [1, 2]. Such BEC interferometry is based on having almost
all bosons in one (or perhaps two) modes.

A typical double-well interferometry experiment involves starting with a BEC
in a single well trap, then changing the trap to a possibly asymmetric
double-well and then back to a single well. Asymmetry could lead to
excitation of bosons to higher energy states of final trap, or to changes to
spatial interference patterns. The process of exciting one boson from the
ground to the first excited state for a single well trap via two quantum
pathways is shown, both pathways involving an intermediate double well trap.
The near degeneracy of energy levels for the intermediate asymmetric double
well facilitates the boson transfer to the excited state. The two
non-observed quantum pathways involve the boson transfer occurring in
different halves of the process, and superposition of the two quantum
transition amplitudes can lead to interference effects in the excitation
probability.

Interferometry experiments involving measurements of boson positions are
best described in terms of quantum correlation functions, which are
expectation values of products of bosonic field operators specified at
different spatial points, and are related to many-boson position
measurements [3]. If boson-boson interactions were absent and the BEC
isolated from the environment, idealised forms of quantum correlation
functions would result, with interferometric effects clearly visible.
Interactions of the BEC with the external environment (such as fluctuating
trap fields) and internal boson-boson interactions tend to degrade the
interference pattern. Boson-boson interactions can result in dephasing
(associated with interactions within condensate modes) and decoherence
effects (associated with interactions causing transitions from condensate
modes) even when external environmental effects are absent.

Recently a simple theory of double-well BEC interferometry has been
developed [4] based on a two mode approximation and allowing for possible
fragmentations of the original BEC into two modes - which may be localised
in each well. Many previous theories ignore fragmentation, with all bosons
occupying a single condensate wave function which satisfies the standard
Gross-Pitaevskii equation (see [5] and references therein). The two-mode
theory was developed from the quantum principle of least action and gives
self consistent coupled equations for the two mode functions (generalised
Gross-Pitaevskii equations) and for the amplitudes (matrix equations)
describing fragmentation of the BEC. Self consistency results in the mode
functions depending on the relative importance of various ways the BEC can
fragment, whilst the fragmentation amplitudes depend on the mode functions.
Numerical studies are planned. However, only transitions within the two
condensate modes are allowed for, and although some dephasing processes are
included, decoherence processes associated with elementary collective
excitations (Bogoliubov) and single boson excitations (thermal modes) are
not. The theory is also restricted to small $N$.

A full treatment of decoherence and dephasing effects on quantum correlation
functions is required, and this is the subject of the present paper. Several
possible approaches to developing such a theory may be identified. These all
have desirable features such as exploiting the physics of large occupancy
differences for condensate and non-condensate modes, not being restricted to
small boson numbers and avoiding the explicit consideration of large numbers
of modes. Collective and single boson excitations from condensate modes, and
possible fragmentation effects can be allowed for, though the presence of
these processes may not be explicit.

One such approach is a master equation method\ [6], in which a condensate
density operator is defined for which a master equation is derived allowing
for interactions with non-condensate modes, which constitute a reservoir.
The difficulty with this method is that it is hard to evaluate the
non-condensate contributions to quantum correlation functions. A second
approach is based on a Heisenberg equation method\ that has been applied in
numerous many-body theory cases. Heisenberg equations for field operators
and products of field operators are derived, and taking the expectation
values with the initial density operator results in a heirarchy of coupled
equations for quantum correlation functions. An ansatz (such as assuming
that a suitable high order correlation function factorises) produces a
truncated set of coupled equations from which correlation functions of the
required order can be calculated. The problem with this method is that it is
hard to confirm the validity of the ansatz.

The present approach is a generalised phase space method, but involving a
distribution functional rather than a distribution function [7]. The field
operator for the bosonic system is written as a sum of condensate and
non-condensate mode contributions [8]. The BEC\ state is described by a
density operator which satisfies the Liouville-von Neumann equation (LVN),
and which is mapped onto a phase space distribution functional. The latter
has the feature that the highly occupied condensate modes are described via
a generalised Wigner representation (since the bosons in condensate modes
behave like a classical mean field), whilst the basically unoccupied
non-condensate modes are described via a positive P representation (these
bosons should exhibit quantum effects). The LVN equation is replaced by a
functional Fokker-Planck equation (FFPE) for the distribution functional,
which is based on the truncated Wigner approximation [7] that can be applied
when large condensate mode occupancy occurs. The FFPE are finally replaced
by coupled Ito stochastic equations (c-number Langevin equations) for
condensate and non-condensate field functions, where the Ito equations
contain deterministic and random noise terms - identifiable from the FFPE.
Stochastic averages of the field functions then give the quantum correlation
functions. There are no obvious difficulties with this approach, though
further development will be needed to explicitly incorporate Bogoliubov
collective excitations.\bigskip

\section{Generalised phase space functional theory}

The Hamiltonian in terms of bosonic field operators $\widehat{\Psi }(\mathbf{%
r}),\widehat{\Psi }(\mathbf{r})^{\dag }$ is%
\begin{eqnarray}
\widehat{H} &=&\dint d\mathbf{r(}\frac{\hbar ^{2}}{2m}\nabla \widehat{\Psi }(%
\mathbf{r})^{\dag }\cdot \nabla \widehat{\Psi }(\mathbf{r})+\widehat{\Psi }(%
\mathbf{r})^{\dag }V\widehat{\Psi }(\mathbf{r}))  \nonumber \\
&&+\dint d\mathbf{r}\frac{g}{2}\widehat{\Psi }(\mathbf{r})^{\dag }\widehat{%
\Psi }(\mathbf{r})^{\dag }\widehat{\Psi }(\mathbf{r})\widehat{\Psi }(\mathbf{%
r}),  \label{Eq. Hamiltonian}
\end{eqnarray}%
where the boson mass is $m$, the trap potential is $V(\mathbf{r},t)$ and $g$
specifies boson-boson interactions in the zero range approximation.

The field operator $\widehat{\Psi }(\mathbf{r})$ is written as sum of a
condensate term $\widehat{\Psi }_{C}(\mathbf{r})$ and a non-condensate term $%
\widehat{\Psi }_{NC}(\mathbf{r})$ [6]%
\begin{equation}
\widehat{\Psi }_{C}(\mathbf{r})=\widehat{c}_{1}\phi _{1}(\mathbf{r})+%
\widehat{c}_{2}\phi _{2}(\mathbf{r})\qquad \widehat{\Psi }_{NC}(\mathbf{r}%
)=\dsum\limits_{k\neq 1,2}^{n}\widehat{c}_{k}\phi _{k}(\mathbf{r}),
\label{Eq.Cond&NonCondField}
\end{equation}%
which are defined in terms of $n$ orthonormal mode functions $\phi _{k}(%
\mathbf{r})$ and bosonic mode annihilation operators $\widehat{c}_{k}$.
Condensate and non-condensate contributions to the field operators commute.

Replacing $\widehat{\Psi }(\mathbf{r})$ by $\widehat{\Psi }_{C}(\mathbf{r})+%
\widehat{\Psi }_{NC}(\mathbf{r})$ the Hamiltonian is sum of three terms [6],
Hamiltonians $\widehat{H}_{C}$ and $\widehat{H}_{NC}$ for the condensate and
non-condensate - which are of the same form as in Equation (\ref{Eq.
Hamiltonian}), and the interaction $\widehat{V}$ between condensate and
non-condensate. This is the sum of three contributions, which are linear,
quadratic and cubic in the condensate operators.

Spatial coherence effects for interference experiments in BECs may be
described via quantum correlation functions, which depend on the density
operator $\widehat{\rho }$ for the bosonic system and are defined by%
\begin{eqnarray}
&&G^{N}(\mathbf{r}_{1}\mathbf{,r}_{2}\mathbf{,..,r}_{N};\mathbf{s}_{N}%
\mathbf{,..,s}_{2}\mathbf{,s}_{1})  \nonumber \\
&=&Tr(\widehat{\rho }(t)\,\widehat{\Psi }\,(\mathbf{r}_{1})^{\dag }\,..%
\widehat{\Psi }\,(\mathbf{r}_{N})^{\dag }\,\widehat{\Psi }\,(\mathbf{s}%
_{N})\,..\,\widehat{\Psi }\,(\mathbf{s}_{1})).  \label{Eq.QuantumCorrFns}
\end{eqnarray}%
The quantum correlation function with $r_{i}=s_{i}\,(i=1,...,N)$ determines
the simultaneous probability of detecting one boson at $r_{1}$, .., the $N$%
th at $r_{N}$, see [3]. It is evident that the quantum correlation functions
will contain condensate terms (describing the main interference effects),
non-condensate terms and mixed terms involving both condensate and
non-condensate operators (describing effects degrading the interference
patterns).

In the phase space functional method the density operator $\widehat{\rho }$
is first mapped uniquely onto a characteristic functional $\chi \lbrack \xi
_{C}(\mathbf{r}),\xi _{C}^{+}(\mathbf{r}),\xi _{NC}(\mathbf{r}),\xi
_{NC}^{+}(\mathbf{r})]$ of the four functions $\xi _{C}^{+}(\mathbf{r}),\xi
_{C}(\mathbf{r}),\xi _{NC}^{+}(\mathbf{r})$ and $\xi _{NC}(\mathbf{r})$%
\begin{eqnarray}
&&\chi \lbrack \xi _{C},\xi _{C}^{+},\xi _{NC},\xi _{NC}^{+}]  \nonumber \\
&=&Tr(\widehat{\rho }\,\exp i\dint d\mathbf{r}\{\xi _{C}(\mathbf{r})\widehat{%
\Psi }_{C}^{\dag }(\mathbf{r})+\widehat{\Psi }_{C}(\mathbf{r})\xi _{C}^{+}(%
\mathbf{r})\}\,  \nonumber \\
&&\times \exp i\dint d\mathbf{r}\{\xi _{NC}(\mathbf{r})\widehat{\Psi }%
_{NC}^{\dag }(\mathbf{r})\}\,\exp i\dint d\mathbf{r}\{\widehat{\Psi }_{NC}(%
\mathbf{r})\xi _{NC}^{+}(\mathbf{r})\})  \label{Eq.CharFnal}
\end{eqnarray}%
The characteristic functional is of the Wigner $(W)$ type for condensate
modes and the positive P\emph{\ }$(P^{+})$ type for the non-condensate modes.

The (quasi) distribution functional $P[\psi _{C}(\mathbf{r}),\psi _{C}^{+}(%
\mathbf{r}),\psi _{NC}(\mathbf{r}),\psi _{NC}^{+}(\mathbf{r})]$ involves
four field functions $\psi _{C}(\mathbf{r}),\psi _{C}^{+}(\mathbf{r}),\psi
_{NC}(\mathbf{r}),\psi _{NC}^{+}(\mathbf{r})$ corresponding to field
operators $\widehat{\Psi }_{C}(\mathbf{r}),\widehat{\Psi }_{C}(\mathbf{r}%
)^{\dag },\widehat{\Psi }_{NC}(\mathbf{r})$ and $\widehat{\Psi }_{NC}(%
\mathbf{r})^{\dag }$. Although non-unique and possibly negative (and hence
not interpretable as a probability distribution), it is required to
determine the characteristic functional $\chi \lbrack \xi _{C}(\mathbf{r}%
),\xi _{C}^{+}(\mathbf{r}),\xi _{NC}(\mathbf{r}),\xi _{NC}^{+}(\mathbf{r})]$
via a functional integration process with weight function $w(\psi _{1},\psi
_{1}^{+},..,\psi _{i},\psi _{i}^{+},..,\psi _{n},\psi _{n}^{+},)$ given by $%
\dprod\limits_{i}(\Delta \mathbf{r}_{i})$ 
\begin{eqnarray}
&&\chi \lbrack \xi _{C}(\mathbf{r}),\xi _{C}^{+}(\mathbf{r}),\xi _{NC}(%
\mathbf{r}),\xi _{NC}^{+}(\mathbf{r})]  \nonumber \\
&=&\diiiint D^{2}\psi _{C}\,D^{2}\psi _{C}^{+}\,D^{2}\psi _{NC}\,D^{2}\psi
_{NC}^{+}\,\,P[\psi _{C}(\mathbf{r}),\psi _{C}^{+}(\mathbf{r}),\psi _{NC}(%
\mathbf{r}),\psi _{NC}^{+}(\mathbf{r})]  \nonumber \\
&&\times \exp i\dint d\mathbf{r\,\{}\xi _{C}(\mathbf{r})\psi _{C}^{+}(%
\mathbf{r})+\psi _{C}(\mathbf{r})\xi _{C}^{+}(\mathbf{r})\}\,
\label{Eq.DistribnFnal} \\
&&\times \exp i\dint d\mathbf{r\,\{}\xi _{NC}(\mathbf{r})\psi _{NC}^{+}(%
\mathbf{r})\}\exp i\dint d\mathbf{r\,\{}\psi _{NC}(\mathbf{r})\xi _{NC}^{+}(%
\mathbf{r})\}.  \nonumber
\end{eqnarray}

Quantum averages of symmetrically ordered products of condensate field
operators $\{\widehat{\Psi }_{C}^{\dag }(\mathbf{r}_{1})....\widehat{\Psi }%
_{C}^{\dag }(\mathbf{r}_{p})\widehat{\Psi }_{C}(\mathbf{s}_{q})..\widehat{%
\Psi }_{C}(\mathbf{s}_{1})\}$ and normally ordered products of
non-condensate field operators $\widehat{\Psi }_{NC}^{\dag }(\mathbf{u}_{1})%
\widehat{\Psi }_{NC}^{\dag }(\mathbf{u}_{2})....\widehat{\Psi }_{NC}^{\dag }(%
\mathbf{u}_{r})\widehat{\Psi }_{NC}(\mathbf{v}_{s})..\widehat{\Psi }_{NC}(%
\mathbf{v}_{1})$ are given by functional integrals of the distribution
functional $P[\psi _{C},\psi _{C}^{+},\psi _{NC},\psi _{NC}^{+}]$ with
products of field functions,.where the condensate field operator $\widehat{%
\Psi }_{C}(\mathbf{r}_{i})^{\dag }$ is replaced by $\psi _{C}^{+}(\mathbf{r}%
_{i})$, $\widehat{\Psi }_{C}(\mathbf{s}_{j})$ is replaced by $\psi (\mathbf{s%
}_{j})$ and with analogous replacements for the non-condensate field
operators.%
\begin{eqnarray}
&&Tr[\widehat{\rho }\,\{\widehat{\Psi }_{C}^{\dag }(\mathbf{r}_{1})...%
\widehat{\Psi }_{C}^{\dag }(\mathbf{r}_{p})\widehat{\Psi }_{C}(\mathbf{s}%
_{q})..\widehat{\Psi }_{C}(\mathbf{s}_{1})\}\,  \label{Eq.QuantumAverages} \\
&&\times \widehat{\Psi }_{NC}^{\dag }(\mathbf{u}_{1})...\widehat{\Psi }%
_{NC}^{\dag }(\mathbf{u}_{r})\widehat{\Psi }_{NC}(\mathbf{v}_{s})..\widehat{%
\Psi }_{NC}(\mathbf{v}_{1})]  \nonumber \\
&=&\diiiint D^{2}\psi _{C}\,D^{2}\psi _{C}^{+}\,D^{2}\psi _{NC}\,D^{2}\psi
_{NC}^{+}\,\,P[\psi _{C}(\mathbf{r}),\psi _{C}^{+}(\mathbf{r}),\psi _{NC}(%
\mathbf{r}),\psi _{NC}^{+}(\mathbf{r})]  \nonumber \\
&&\times \psi _{C}^{+}(\mathbf{r}_{1})\,..\psi _{C}^{+}(\mathbf{r}%
_{p})\,\psi _{C}(\mathbf{s}_{q})\,...\psi _{C}(\mathbf{s}_{1})\,\psi
_{NC}^{+}(\mathbf{u}_{1})\,..\psi _{NC}^{+}(\mathbf{u}_{r})\,\psi _{NC}(%
\mathbf{v}_{s})...\psi _{NC}(\mathbf{v}_{1}).  \nonumber
\end{eqnarray}%
Symmetric ordering is defined as the average over all $N(p,q)=(p+q)!$
permutations of the factors $\widehat{\Psi }^{\dag }(\mathbf{r}_{1})..%
\widehat{\Psi }^{\dag }(\mathbf{r}_{p})\widehat{\Psi }(\mathbf{s}_{q})..%
\widehat{\Psi }(\mathbf{s}_{1})$. These results plus equal time commutation
rules give the quantum correlation functions.

For example, the first order quantum correlation function that is used to
exhibit macroscopic spatial coherence in BECs is given by%
\begin{eqnarray}
&&G^{1}(\mathbf{r}_{1};\mathbf{s}_{1})=\left\langle \,\widehat{\Psi }(%
\mathbf{r}_{1})^{\dag }\,\widehat{\Psi }(\mathbf{s}_{1})\right\rangle 
\nonumber \\
&=&-\frac{1}{2}\delta (\mathbf{r}_{1}-\mathbf{s}_{1})
\label{Eq.FirstOrderCorrFn} \\
&&+\diiiint D^{2}\psi _{C}\,D^{2}\psi _{C}^{+}\,D^{2}\psi _{NC}\,D^{2}\psi
_{NC}^{+}\,\,P[\psi _{C}(\mathbf{r}),\psi _{C}^{+}(\mathbf{r}),\psi _{NC}(%
\mathbf{r}),\psi _{NC}^{+}(\mathbf{r})]  \nonumber \\
&&\times (\psi _{C}^{+}(\mathbf{r}_{1})+\,\psi _{NC}^{+}(\mathbf{r}%
_{1}))(\psi _{C}(\mathbf{s}_{1})+\,\psi _{NC}(\mathbf{s}_{1})).  \nonumber
\end{eqnarray}%
The result includes pure condensate terms, pure non-condensate terms and
mixed terms. The delta function term arises from the difference between
normal and symmetric ordering for the condensate terms.

The Liouville-von Neumann equation for the density operator replaced by the
functional Fokker-Planck equation for the distribution functional by using
the correspondence rules involving functional derivatives, such as $\widehat{%
\Psi }_{C}(\mathbf{s})\widehat{\rho }\leftrightarrow \left( \psi _{C}(%
\mathbf{s})+\frac{1}{2}\delta /\delta \psi _{C}^{+}(\mathbf{s})\right)
P[\psi _{C},..,\psi _{NC}^{+}]$ and $\widehat{\Psi }_{NC}(\mathbf{s})%
\widehat{\rho }\leftrightarrow \left( \psi _{NC}(\mathbf{s})\right) P[\psi
_{C},..,\psi _{NC}^{+}]$. The density operator is multiplied by various
field operators in the LVN equation and the overall effect on the
distribution functional is obtained by applying such rules in succession.
The general form for the FFPE after applying the truncated Wigner
approximation to remove the third order functional derivatives is%
\begin{eqnarray}
&&\frac{\partial }{\partial t}P[\psi _{C},\psi _{C}^{+},\psi _{NC},\psi
_{NC}^{+}]  \nonumber \\
&=&\dint d\mathbf{r\{}\dsum\limits_{\alpha =1}^{4}-\frac{\delta }{\delta
\phi _{\alpha }}A_{\alpha }(\psi _{C},\psi _{C}^{+},\psi _{NC},\psi
_{NC}^{+})  \nonumber \\
&&+\frac{1}{2}\dsum\limits_{\alpha ,\beta =1}^{4}\frac{\delta ^{2}}{\delta
\phi _{\alpha }\delta \phi _{\beta }}D_{\alpha \beta }(\psi _{C},\psi
_{C}^{+},\psi _{NC},\psi _{NC}^{+})\}  \nonumber \\
&&\times P[\psi _{C},\psi _{C}^{+},\psi _{NC},\psi _{NC}^{+}],
\label{Eq.FFPE}
\end{eqnarray}%
where $\phi _{1}\equiv \psi _{C},\phi _{2}\equiv \psi _{C}^{+},\phi
_{3}\equiv \psi _{NC},\phi _{4}\equiv \psi _{NC}^{+}$. The drift vector $%
A_{\alpha }$ involves the fields $\psi _{C},\psi _{C}^{+},\psi _{NC},\psi
_{NC}^{+}$ and their spatial derivatives. The positive definite diffusion
matrix $D_{\alpha \beta }$ also involves these fields.

The functional Fokker-Planck equation for the distribution functional is
then replaced by Ito stochastic field equations for $\psi _{C},\psi
_{C}^{+},\psi _{NC},\psi _{NC}^{+}$, which are now regarded as time
dependent stochastic fields. The general form for the stochastic field
equations is%
\begin{equation}
\frac{\partial }{\partial t}\phi _{\alpha }=A_{\alpha }(\psi _{C},\psi
_{C}^{+},\psi _{NC},\psi _{NC}^{+})+\dsum\limits_{\beta }d_{\alpha \beta
}(\psi _{C},\psi _{C}^{+},\psi _{NC},\psi _{NC}^{+})\Gamma _{\beta }(\mathbf{%
r},t).  \label{Eq.ItoStochastic}
\end{equation}%
The first term is deterministic and is associated with the drift vector. The
second term is a random noise term and is associated with diffusion matrix,
the matrix $d$ being related to the diffusion matrix via $D=d\,d^{T}$. The $%
\Gamma _{\beta }(\mathbf{r},t)$ are Gaussian-Markov random noise terms%
\begin{equation}
\overline{\Gamma _{\alpha }(\mathbf{r}_{1},t_{1})}=0\qquad \overline{\Gamma
_{\alpha }(\mathbf{r}_{1},t_{1})\Gamma _{\beta }(\mathbf{r}_{2},t_{2})}%
=\delta _{\alpha \beta }\delta (t_{1}-t_{2})\delta (\mathbf{r}_{1}-\mathbf{r}%
_{2})  \label{Eq.GauusianMarkoff}
\end{equation}%
where the bar denotes a stochastic average.

Quantum averages of the field operators are now given by stochastic
averages, which are equivalent to the functional integrals in Equation (\ref%
{Eq.QuantumAverages}) involving the distribution functional.

\begin{eqnarray}
&&Tr[\widehat{\rho }\,\{\widehat{\Psi }^{\dag }(\mathbf{r}_{1})..\widehat{%
\Psi }^{\dag }(\mathbf{r}_{p})\widehat{\Psi }(\mathbf{s}_{q})..\widehat{\Psi 
}(\mathbf{s}_{1})\}\,\widehat{\Psi }_{NC}^{\dag }(\mathbf{u}_{1})..\widehat{%
\Psi }_{NC}^{\dag }(\mathbf{u}_{r})\widehat{\Psi }_{NC}(\mathbf{v}_{s})..%
\widehat{\Psi }_{NC}(\mathbf{v}_{1})]  \nonumber \\
&=&\overline{\psi _{C}^{+}(\mathbf{r}_{1})\,..\psi _{C}^{+}(\mathbf{r}%
_{p})\,\psi _{C}(\mathbf{s}_{q})\,..\psi _{C}(\mathbf{s}_{1})\,\psi
_{NC}^{+}(\mathbf{u}_{1})\,..\psi _{NC}^{+}(\mathbf{u}_{r})\,\psi _{NC}(%
\mathbf{v}_{s})..\psi _{NC}(\mathbf{v}_{1})}  \nonumber \\
&&  \label{Eq.StochAver}
\end{eqnarray}%
\bigskip

\section{Conclusion}

It is shown how the quantum correlation functions required for describing
interferometry using BECs can obtained via stochastic averages of products
of field functions, which satisfy Ito equations derived from the functional
Fokker-Planck equation for the phase space distribution functional that
represents the quantum density operator. The phase space distribution
functional is of the Wigner type for the condensate modes and the positive P
type for the non-condensate modes. Decoherence and dephasing effects in BEC
interferometry with large boson numbers are fully treated, unlike previous
theories in which only condensate modes are considered or the theory is
restricted to small boson numbers.\bigskip

\section{Acknowledgements}

The author is grateful for discussions with M K Olsen and P D Drummond on
applying Wigner and positive P phase space distributions to BEC\
interferometry.

\end{document}